\documentclass{ws-procs9x6}

\def\etal {{\it et al.}}

\def\ga{\gamma}
\def\de{\delta}
\def\ep{\epsilon}

\def\ka{\kappa}
\def\la{\lambda}

\def\cl{{\cal L}}

\newcommand{\beq}{\begin{equation}}
\newcommand{\eeq}{\end{equation}}
\newcommand{\bea}{\begin{eqnarray}}
\newcommand{\eea}{\end{eqnarray}}

\def\kf{\hat k_F}
\def\kaf{\hat k_{AF}}

\def\voc{\mathrel{\rlap{\lower0pt\hbox{\hskip1pt{$c$}}}
    \raise3pt\hbox{$\neg$}}}
\def\vok{\mathrel{\rlap{\lower0pt\hbox{\hskip1pt{$k$}}}
    \raise6pt\hbox{$\neg$}}}

\def\sk#1#2#3{#1^{(#2)}_{#3}}

\def\kjm#1#2#3{k^{(#1)}_{(#2)#3}}
\def\cjm#1#2#3{c^{(#1)}_{(#2)#3}}
\def\kI{\cjm{d}{I}{jm}}

\def\kE{\kjm{d}{E}{jm}}
\def\kB{\kjm{d}{B}{jm}}
\def\kV{\kjm{d}{V}{jm}}

\def\cftzE{\sk{(\voc_F^{(d)})}{0E}{njm}}

\def\ae{\widehat a_{\rm eff}}
\def\ce{\widehat c_{\rm eff}}
\def\He{\widehat H_{\rm eff}}
\def\ge{\widehat g_{\rm eff}}
\def\nuTemplate#1#2#3#4{\big(#1^{(#2)}\big)_{#3}^{#4}}
\def\ceff#1#2#3{\nuTemplate{c_\text{eff}}{#1}{#2}{#3}}
\def\aeff#1#2#3{\nuTemplate{a_\text{eff}}{#1}{#2}{#3}}
\def\Heff#1#2#3{\nuTemplate{H_\text{eff}}{#1}{#2}{#3}}
\def\geff#1#2#3{\nuTemplate{g_\text{eff}}{#1}{#2}{#3}}
\def\cof#1#2{\nuTemplate{c_\text{of}}{#1}{#2}{}}
\def\aof#1#2{\nuTemplate{a_\text{of}}{#1}{#2}{}}

\begin{document}

\title{HIGHER-ORDER LORENTZ VIOLATION}

\author{MATTHEW MEWES}

\address{Department of Physics and Astronomy,
  Swarthmore College\\
  Swarthmore, PA 19081, USA}

\begin{abstract}
  This brief review discusses
  Lorentz-violating operators
  of arbitrary dimension
  within the photon and neutrino sectors
  of the Standard-Model Extension.
\end{abstract}

\bodymatter

\section{Introduction}

The Standard-Model Extension (SME) provides a general
framework for theoretical and experimental
studies of Lorentz and CPT violation.\cite{datatables}
Most early research focused
on the minimal SME (mSME),
which restricts attention to operators of 
renormalizable dimensions $d=3$ and $4$ 
in flat spacetime.\cite{ck}
However, the full SME encompasses
curved spacetime\cite{smegrav}
and includes operators of arbitrary dimension.
Here, we give a brief discussion
of recent efforts to classify terms
in the photon and neutrino
sectors of the SME,
including those of nonrenormalizable
dimensions $d\geq 5$.
Details can be found
in Refs.\ \refcite{km09} and \refcite{km12}.
A summary of experimental results
is given in Ref.\ \refcite{datatables}.

\section{Photons}

The pure photon sector of the SME 
is given by the lagrangian\cite{km09}
\beq
\cl =  -\tfrac 1 4 F_{\mu\nu}F^{\mu\nu}
+\tfrac 1 2 \ep^{\ka\la\mu\nu}A_\la (\kaf)_\ka F_{\mu\nu}
- \tfrac 1 4 F_{\ka\la} (\kf)^{\ka\la\mu\nu} F_{\mu\nu} .
\label{lagrangian}
\eeq
In addition to the usual Maxwell term,
there are two Lorentz-violating terms
involving the CPT-odd four-vector $(\kaf)^\ka$
and CPT-even tensor $(\kf)^{\ka\la\mu\nu}$.
Both $(\kaf)^\ka$ and $(\kf)^{\ka\la\mu\nu}$
are constants in the mSME
but are momentum dependent in the full SME,
where they each take the form of a power series
in photon momentum. 
The expansion constants in the series
give coefficients for Lorentz violation.

The total number of coefficients
for Lorentz violation
that appear in the expansions of
$(\kaf)^\ka$ and $(\kf)^{\ka\la\mu\nu}$
grows rapidly ($\sim d^3$)
as we consider higher dimensions.
To aid in classifying the numerous coefficients,
a spherical-harmonic expansion of
$(\kaf)^\ka$ and $(\kf)^{\ka\la\mu\nu}$ is performed,
giving sets of spherical coefficients
for Lorentz violation that can be tested
by different types of experiment.
Several classes of experiment provide
high sensitivity to various subsets of the coefficients,
including
searches for astrophysical birefringence
and dispersion,
and resonant-cavity tests.
Below is a summary of coefficients
that have been experimentally tested thus far.

{\em Birefringent vacuum coefficients.}
The coefficients $\kE$, $\kB$, and $\kV$
lead to birefringence of light propagating {\it in vacuo}.
The result is a change in polarization as light propagates.
The effect can depend on the direction of propagation
and photon energy.
Polarimetry of radiation
from distant astrophysical sources
has led to many constraints
on the mSME coefficients\cite{bire_renorm,kmapjl}
and nonminimal coefficients up to dimension $d=9$.\cite{km09,bire_nonrenorm}

{\em Nonbirefringent vacuum coefficients.}
The coefficients $\kI$ affect the propagation of
light in a polarization-independent way,
implying no birefringence.
However, they do lead to vacuum dispersion for $d\geq 6$.
Time-of-flight tests involving high-energy sources,
such as $\ga$-ray bursts,
search for differences in arrival times of photons
at different energies.
Again, this effect can depend on the photon energy
and direction of propagation.
Constraints on the $\kI$ coefficients
for $d=6$ and $8$ have been found.\cite{km09,kmapjl,fermi}

{\em Camouflage coefficients.}
A large class of Lorentz violation has
no effect on the propagation of light.
These violations are referred
to as vacuum orthogonal.
Among them are the so-called camouflage violations,
which generically give polarization-independent effects,
in addition to giving conventional light propagation.
As a result, the effects of the associated
camouflage coefficients $\cftzE$
are particularly subtle.
They can, however, be tested
in resonant-cavity experiments,
where they lead to tiny shifts
in resonant frequencies.\cite{km09,cavity_theory}
Numerous cavity searches for $d=4$ coefficients
have been performed.\cite{cavity_expt}
A recent experiment placed the first cavity
bounds on nonminimal $d=6$ and $8$
camouflage coefficients.\cite{parker}

\section{Neutrinos}

Neutrinos in the SME are governed
by a $6\times 6$ effective hamiltonian
that acts on the six-dimensional
space that includes both neutrinos and antineutrinos.
The Lorentz-violating part of the hamiltonian
takes the form\cite{km12}
\beq
\de h_\text{eff} =
\frac{1}{|\vec p|} \begin{pmatrix}
  \ae - \ce & -\ge + \He\\
  -\ge^\dag + \He^\dag & -\ae^T - \ce^T
\end{pmatrix} \ .
\eeq
The $3\times 3$ matrices 
$\ae$, $\ce$, $\ge$, and $\He$
are functions
of the neutrino momentum vector $\vec p$,
giving rise to unusual direction dependence
and an unconventional energy-momentum relation.
Note that the $\ge$, and $\He$ terms
cause mixing between neutrinos and antineutrinos.
The $\ce$ and $\He$ matrices contain the CPT-even violations,
while the CPT-odd terms are in $\ae$ and $\ge$.
As with photons,
a spherical-harmonic expansion
is used to enumerate and classify
the various effects and coefficients.
For example,
the $\ae$ matrix has the expansion
$(\ae)^{ab} = \sum |\vec p|^{d-2}\, Y_{jm}(\hat p) \aeff{d}{jm}{ab}$,
giving spherical coefficients for Lorentz violation 
$\aeff{d}{jm}{ab}$.
Most research in neutrinos, so far,
focuses on one of the following cases.

{\em Oscillation coefficients.}
Except for a small subset
of flavor-diagonal coefficients,
most of the spherical coefficients,
$\aeff{d}{jm}{ab}$,
$\ceff{d}{jm}{ab}$,
$\geff{d}{jm}{ab}$, and
$\Heff{d}{jm}{ab}$,
produce oscillations.
Signatures of Lorentz violation
in oscillations include
direction dependence,
unconventional energy dependence,
neutrino-antineutrino oscillations,
and CPT asymmetries.
Neutrino oscillations are interferometric in nature,
so high sensitivity to Lorentz violation is possible.
While some bounds on coefficients up to $d=10$
have been deduced from earlier analyses,\cite{km12}
most of the research so far has
focused on the $d=3$ and $4$ cases,\cite{nu_theory}
with many constraints
on mSME coefficients.\cite{nu_expt}

{\em Oscillation-free coefficients.}
Lorentz violation also affects the
kinematics of neutrino propagation.
Kinematic effects are characterized,
independent of oscillations,
in the simple oscillation-free limit,
where oscillations are neglected,
and all neutrinos are treated the same.
The energy in this limit is given by
\beq
E = |\vec p| + \tfrac{|m_l|^2}{2|\vec p|}
+ \sum |\vec p|^{d-3} Y_{jm}(\hat p) \Big[\aof{d}{jm}-\cof{d}{jm}\Big]\ ,
\eeq
where $\aof{d}{jm}$ and $\cof{d}{jm}$
are oscillation-free coefficients.
This provides a framework
for a range of studies,
such as time-of-flight tests,
analyses of meson-decay thresholds,
and \v Cerenkov-like decays of neutrinos,
each of which has produced constraints
on coefficients up to $d=10$.\cite{km12,alt}

\end{document}